\documentclass[final,twocolumn]{book}

\usepackage[english]{babel}
\usepackage{lipsum}

\usepackage{amsfonts}
\usepackage{dcolumn}
\usepackage{bm}
\usepackage{amsmath}
\usepackage{nicefrac}
\usepackage{amsthm}
\usepackage{amssymb}
\usepackage{graphicx}
\usepackage{color}
\usepackage{url}
\usepackage{cancel}
\usepackage{rotating}

\usepackage{pdfpages}

\usepackage[colorlinks=true,linkcolor=linkcolour,citecolor=linkcolour,urlcolor=linkcolour,bookmarksopen]{hyperref}
\definecolor{linkcolour}{rgb}{0.0,0.0,0.55} 

\newcommand{\appropto}{\mathrel{\vcenter{
  \offinterlineskip\halign{\hfil$##$\cr
    \propto\cr\noalign{\kern2pt}\sim\cr\noalign{\kern-2pt}}}}}
\renewcommand{\v}[1]{\boldsymbol{#1}}		

\usepackage{float}
\usepackage{feynmp}
\usepackage{enumerate}
\numberwithin{equation}{section}
\usepackage{makeidx}
\makeindex

\begin{document}


\setcounter{part}{1}
\setcounter{chapter}{30}

\chapter[Searches for New Particles Including Dark Matter]{Searches for New Particles Including Dark Matter with Atomic, Molecular and Optical Systems}
\label{Chapter:DM}

\section*{\emph{Abstract}}
\emph{
The ``standard model'' of physics has been successful in explaining most physical processes and phenomena that we see around us. 
However, despite the great success of the standard model, there remain a number of unresolved puzzles within the model, as well as questions about the self-consistency of the framework. 
Additionally, various independent astrophysical and cosmological observations contradicting the predictions of the standard model have been accumulating over the course of the past century. 
Many of these puzzles and unexpected observations can be elegantly explained by postulating the existence of at least one new particle or field outside of the present standard model. 
}

\emph{
New particles can manifest their effects in many settings, ranging from effects on sub-atomic to galactic length scales. 
The nature of these effects depends on the specific particles and their non-gravitational interactions. 
In this chapter, we give a brief overview of how atomic, molecular and optical systems can be used to search for new particles. 
To illustrate the basic principles behind these methods, we focus on the simplest class of particles, namely new spinless bosons. 
}

\pagebreak

The ``standard model'' of particle physics at present provides the most fundamental framework for understanding the basic building blocks of matter and describing the various known interactions between these building blocks. 
The standard model does incredibly well in describing physical processes and phenomena that take place over a very broad range of energies and length scales, from explaining the binding of the constituents of atoms, to understanding the formation and evolution of stars. 

\index{dark matter}
However, numerous astrophysical and cosmological observations have been accumulating over the course of the past century that cannot be explained by the standard model. 
Observations of stellar orbits about the galactic centre from as early as the 1930s indicate the presence of a non-baryonic matter component that is traditionally termed ``dark matter'' (this non-baryonic matter component does not appreciably emit or absorb electromagnetic radiation) \cite{DM_Review_2005}. 
Further astrophysical evidence for dark matter comes from measurements of angular fluctuations in the cosmic microwave background spectrum \cite{Planck_2015} and the need for non-baryonic matter to explain the observed structure formation in our Universe \cite{DM_Review_2005}. 
\index{dark energy}
Additionally, distance and redshift measurements of supernovae show that the expansion of the Universe is accelerating, indicating that the Universe is being pushed apart by a repulsive force associated with a ``dark energy'' component \cite{DE_Review_2008}. 
These dark components (which are naturally explained by postulating the existence of at least one new particle or field) overwhelmingly dominate the observed matter-energy content of our Universe, with ordinary baryonic matter making up only a small fraction of the total content \cite{Planck_2015}. 

\index{baryogenesis}
Another profound mystery is the observed predominance of matter over antimatter in our Universe --- the problem of baryogenesis. 
The standard model contains the necessary ingredients to produce ever slightly more matter than antimatter; however, the observed predominance of matter over antimatter in our Universe is much larger than can be facilitated within the standard model \cite{Baryogenesis_Review_1999}. 
One of the key ingredients for baryogenesis is \emph{CP} violation, which is the violation of the product of the charge parity (\emph{C} $=$ exchange of particles and antiparticles) and parity (\emph{P} $=$ inversion of spatial coordinates) symmetries. 
Additional sources of \emph{CP} violation necessary to explain baryogenesis may come from new particles possessing \emph{CP}-violating interactions with ordinary matter. 
\index{strong CP problem}
Intriguingly, practically no \emph{CP} violation has been observed in strong processes in the standard model (compared with the relatively large amount of \emph{CP} violation in weak processes). 
\index{axion}
This puzzling observation --- termed the ``strong CP problem'' --- is most elegantly explained by postulating the existence of a new low-mass feebly-interacting spinless boson called the axion \cite{Axions_Review_2010}. 

In order to corroborate or refute models that claim to explain the above problems and observations via putative new particles, one needs experimental probes for such particles. 
New particles may arise in several different settings:

(1) As mediators of new interactions between particles or bodies (Sec.~\ref{Sec:New_forces}). 

(2) Produced in laboratories or colliders (Sec.~\ref{Sec:Lab_sources}). 

(3) Produced in stars and astrophysical processes (Sec.~\ref{Sec:Astro_sources}). 

(4) Constitute the observed dark matter or dark energy (Sec.~\ref{Sec:Cosmo_sources}).

Atomic, molecular and optical systems lie at the heart of some of the highest precision measurements known to mankind. 
Optical clocks, which measure transition frequencies in atoms and ions, have demonstrated a fractional precision at the level $\sim 10^{-18}$ \cite{Clocks_Review_2011,Clocks_Review_2015}. 
Optical magnetometers, which measure magnetic fields using atoms, have demonstrated a magnetic field sensitivity at the level $\sim 10^{-15}~\textrm{T} ~ \textrm{Hz}^{-1/2}$ \cite{Magnetometers_Review_2007}. 
Laser interferometers (which have directly detected gravitational waves) have demonstrated an equivalent sensitivity to length fluctuations at the level $\sim 10^{-23} ~ \textrm{Hz}^{-1/2}$ \cite{LIGO_2016}. 

Can these extraordinary levels of precision and sensitivity be leveraged to search for new particles? 
The answer to this question is in the affirmative. 
Indeed, new particles arising in all types of settings described above can be sought for with experiments using atomic, molecular and optical systems. 
In this chapter, we present a brief overview of how this can be done, focusing mainly on new spinless bosons (which are the simplest possibility from the theoretical point of view) to help illustrate the basic principles behind the methods. 
We begin by presenting the simplest possible non-gravitational interactions of spinless bosons with ordinary matter (Sec.~\ref{Sec:Interactions}). 
We then explain how atomic, molecular and optical systems can be used to search for spinless bosons possessing non-gravitational interactions in a broad variety of settings (Secs.~\ref{Sec:New_forces}, \ref{Sec:Lab_sources}, \ref{Sec:Astro_sources} and \ref{Sec:Cosmo_sources}). 
Unless explicitly stated otherwise, we adopt the natural units $\hbar = c =1$ in this chapter.

\section{Non-Gravitational Interactions of Spinless Bosons}
\label{Sec:Interactions}
The possible non-gravitational interactions of spinless bosons can be broadly distinguished on the basis of the parity symmetry (behaviour under the inversion of spatial coordinates). 
The most relevant scalar-type (even-parity) interactions of a spinless boson $\phi$ with ordinary matter are:
\begin{equation}
\label{lin_scalar_ints}
\mathcal{L}_\textrm{scalar}^\textrm{lin.} = \frac{g_\gamma^s }{4} \phi F_{\mu \nu} F^{\mu \nu} - \phi \sum_\psi g_\psi^s \bar{\psi} \psi \, ,
\end{equation}
\begin{equation}
\label{quad_scalar_ints}
\mathcal{L}_\textrm{scalar}^\textrm{quad.} = \frac{h_\gamma^s }{4} \phi^2 F_{\mu \nu} F^{\mu \nu} - \phi^2 \sum_\psi h_\psi^s \bar{\psi} \psi  \, ,
\end{equation}
where the first term represents the interaction of the spinless boson with the electromagnetic field tensor $F$, with $\tilde{F}$ the dual field tensor, 
and the second term represents the interaction of the spinless boson with a fermion field $\psi$, with $\bar{\psi} = \psi^\dagger \gamma^0$ the Dirac adjoint. 
Here $g_{\gamma,\psi}^s$ and $h_{\gamma,\psi}^s$ are parameters that determine the relevant non-gravitational interaction strengths. 

The most relevant pseudoscalar-type (odd-parity) interactions of a spinless boson $\phi$ with ordinary matter are:
\begin{align}
\label{axion_ints}
\mathcal{L}_\textrm{pseudoscalar} &= \frac{g_\gamma^p }{4} \phi F_{\mu \nu} \tilde{F}^{\mu \nu} + \frac{g_g^p }{4} \phi G_{\mu \nu} \tilde{G}^{\mu \nu} \notag \\
&- i \phi \sum_\psi g_\psi^p \bar{\psi} \gamma_5 \psi \, ,
\end{align}
where the first term represents the interaction of the spinless boson with the electromagnetic field tensor $F$, 
the second term represents the interaction of the spinless boson with the gluonic field tensor $G$, 
and the third term represents the interaction of the spinless boson with a fermion field $\psi$. 
Here $g_{\gamma,g,\psi}^p$ are parameters that determine the relevant non-gravitational interaction strengths.

\section{New Forces}
\label{Sec:New_forces}
In the presence of the non-gravitational interactions in Eqs.~(\ref{lin_scalar_ints}), (\ref{quad_scalar_ints}) and (\ref{axion_ints}), new forces can be mediated between particles or bodies via the exchange of spinless boson(s). 
The simplest possibility involves the exchange of a single boson between two fermions in the presence of the linear-in-$\phi$ interactions in the last terms of Eqs.~(\ref{lin_scalar_ints}) and (\ref{axion_ints}). 
In this case, there are three distinct potentials that arise from the permutation of the two vertex types. 
In the non-relativistic limit, these potentials take the following form \cite{Moody_NF_1984}: 
\begin{equation}
\label{scalar-scalar_potential}
V_{ss}(\v{r}) =  -  g_1^s g_2^s \frac{e^{-m_\phi r}}{ 4 \pi r}  \, ,
\end{equation}
\begin{equation}
\label{scalar-pseudoscalar_potential}
V_{ps}(\v{r}) = +  g_1^p g_2^s  \v{\sigma}_1 \cdot \hat{\v{r}} \left( \frac{1}{r^2} + \frac{m_\phi}{r} \right) \frac{e^{-m_\phi r}}{8 \pi m_1}  \, ,
\end{equation}
\begin{align}
\label{pseudoscalar-pseudoscalar_potential}
V_{pp}(\v{r}) = -  \frac{g_1^p g_2^p}{4}  \left\{  \v{\sigma}_1 \cdot \v{\sigma }_2 \left[ \frac{1}{r^3} + \frac{m_\phi}{r^2} + \frac{4 \pi}{3} \delta(\v{r}) \right] \right. \notag \\
- \left. \left( \v{\sigma}_1 \cdot \hat{\v{r}} \right) \left( \v{\sigma }_2 \cdot \hat{\v{r}} \right)  \left[ \frac{3}{r^3} + \frac{3m_\phi}{r^2} + \frac{m_\phi^2}{r} \right]   \right\} \frac{e^{-m_\phi r}}{4 \pi m_1 m_2}  \, .
\end{align}
Here, $m_\phi$ is the mass of the exchanged boson, $\v{\sigma}_1$ and $\v{\sigma}_2$ denote the Pauli spin matrix vectors of the two fermions, $\hat{\v{r}}$ is the unit vector directed from fermion 2 to fermion 1, and $r$ is the distance between the two fermions. 
In Eq.~(\ref{scalar-pseudoscalar_potential}), the cross term (obtained by permuting the particle indices $1 \leftrightarrow 2$) is implicit. 

Other relatively common potentials include the potential mediated by the exchange of a single boson between two bodies with non-zero electromagnetic energies in the presence of the linear-in-$\phi$ interaction in the first term of Eq.~(\ref{lin_scalar_ints}) \cite{Leefer_Dy_2016}, the potential mediated by the exchange of a pair of bosons between two fermions in the presence of the quadratic-in-$\phi$ interactions in the last term of Eq.~(\ref{quad_scalar_ints}) \cite{Pospelov_2BE_2008}, and the potential mediated by the exchange of a pair of fermions (including neutrinos) between two particles \cite{Stadnik_2NE_2018}. 
In the limit when the mass of the exchanged particles is small, these potentials scale as $V(r) \propto 1/r$, $V(r) \propto 1/r^3$ and $V(r) \propto 1/r^5$, respectively.

 \begin{figure}[t!]
\includegraphics[width=\columnwidth]{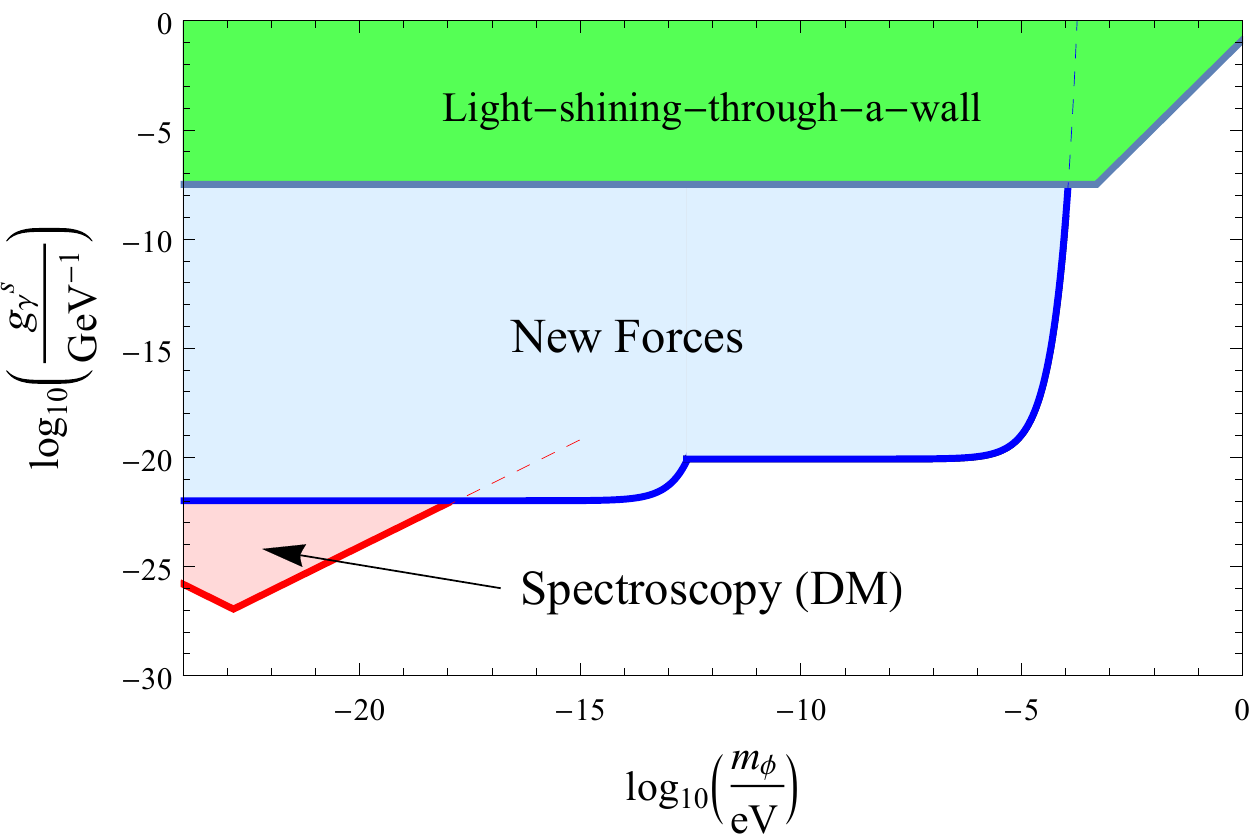}
\caption{Limits on the linear scalar interaction of a spinless boson $\phi$ with the photon, as defined in the first term of Eq.~(\ref{lin_scalar_ints}). 
The region in green corresponds to constraints from light-shining-through-a-wall experiments \cite{ALPS_2010,OSQAR_2015}. 
The region in blue corresponds to constraints from macroscopic-scale experiments that search for new forces \cite{Leefer_Dy_2016,Braginsky_TP_1972,Heckel_TP_1999,Adelberger_TP_2008,Adelberger_TP_2009,Zhan_AI_2015}. 
The region in red corresponds to constraints from atomic spectroscopy measurements that search for the effects of a relic coherently oscillating field $\phi = \phi_0 \cos(m_\phi t)$, which saturates the local cold dark matter (DM) content \cite{Leefer_DM_VFC_2015,Hees_DM_VFC_2016}. 
} 
\label{fig:Lin_scalar_photon}
\end{figure}

\subsection{Macroscopic-scale experiments}
\label{Sec:Macro_scale_exps}
When the condition $m_\phi \ll 1/r$ is satisfied, the potentials in Eqs.~(\ref{scalar-scalar_potential}), (\ref{scalar-pseudoscalar_potential}) and (\ref{pseudoscalar-pseudoscalar_potential}) can be treated as long-range (since the exponential terms reduce to $e^{-m_\phi r} \approx 1$ when $m_\phi r \ll 1$). 
Experiments performed on macroscopic length scales provide an excellent way of probing these new interactions. 
These types of experiments employ a massive body, such as the Sun, Earth, Moon or a massive object in the laboratory, which functions as the source of new bosons. 
In order to detect effects associated with the anomalous interactions mediated by these bosons, a high-precision detector is required. 
Various methods can be used to search for new spin-independent forces in macroscopic-scale experiments: 

\index{torsion pendulum}
(1) Torsion pendula to search for anomalous torques \cite{Braginsky_TP_1972,Heckel_TP_1999,Adelberger_TP_2008,Adelberger_TP_2009}. 

\index{atom interferometry}
(2) Atom interferometers to search for anomalous accelerations \cite{Tino_AI_Review_2014,Zhan_AI_2015}. 

\index{atomic clock}
(3) Atomic clocks and other spectroscopy-based measurements to search for anomalous frequency shifts \cite{Flambaum_NF_2008,Guena_NF_2012,Leefer_Dy_2016}. 
\index{spectroscopy}

The first two types of methods involve measurements of vector quantities, namely differences of torques and accelerations, respectively, of two different test bodies, while the third method involves measuring a scalar quantity, namely the difference in the ratio of two transition frequencies at two different distances from a massive body. 
We mention that lunar laser ranging measurements can also be used to search for spin-independent anomalous interactions \cite{Turyshev_LLR_2004}. 

Various methods can be used to search for new spin-dependent forces in macroscopic-scale experiments: 

\index{torsion pendulum}
(4) Torsion pendula to search for anomalous torques \cite{Hoedl_MD_TP_2011,Terrano_MD_TP_2015}. 

\index{magnetometry}
(5) Magnetometers to search for anomalous spin-precession effects \cite{Youdin_MD_eN_1996,Romalis_NF_2009,Serebrov_MD_NN_2009,Petukhov_MD_NN_2010,Tullney_MD_NN_2013,Snow_MD_NN_2013,Afach_MD_NN_2015,
Ruoso_MD_eN_2017,Rong_MD_eN_2018}. 

\begin{figure}[t!]
\includegraphics[width=\columnwidth]{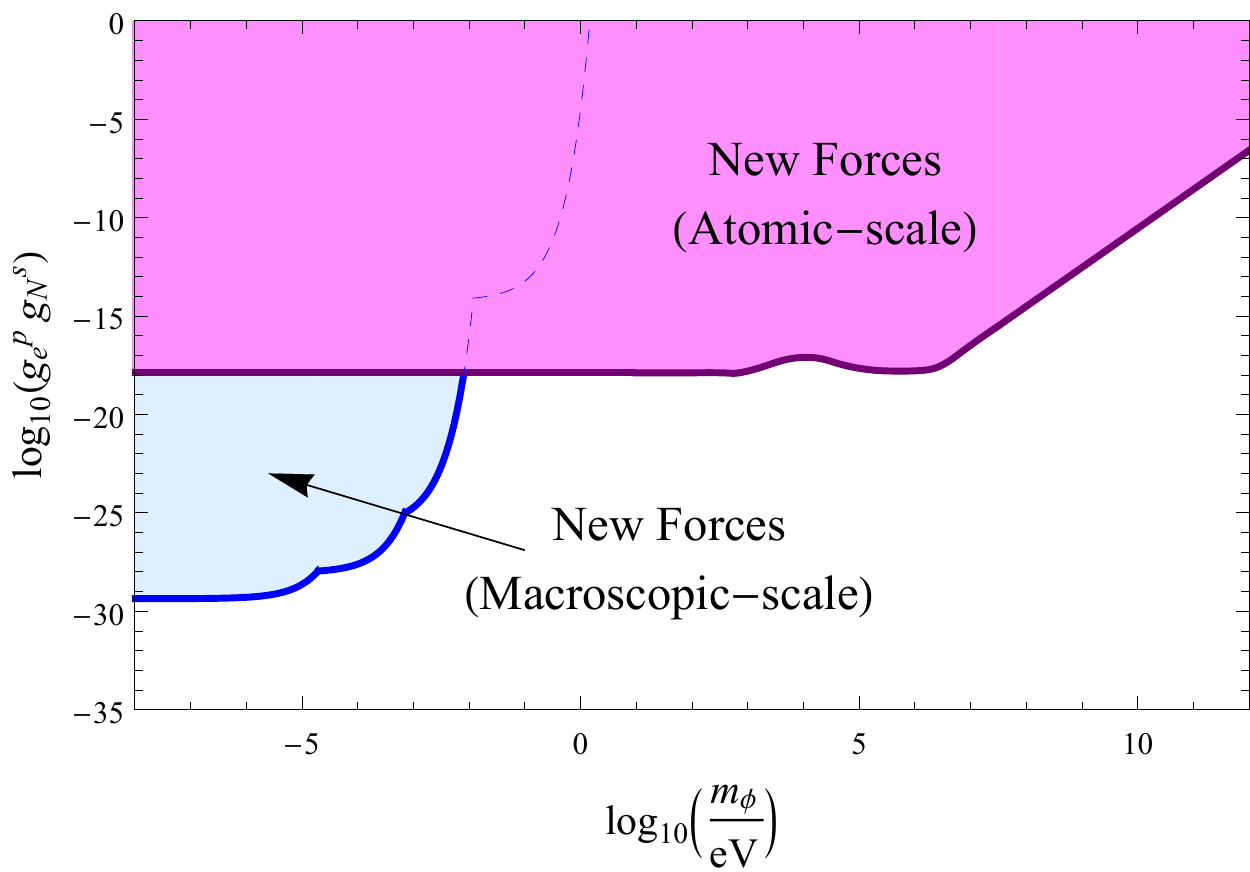}
\caption{Limits on the \emph{P},\emph{T}-violating scalar-pseudoscalar nucleon-electron interaction mediated by a spinless boson $\phi$, as given in Eq.~(\ref{scalar-pseudoscalar_potential}) [which arises from the last terms of Eqs.~(\ref{lin_scalar_ints}) and (\ref{axion_ints})]. 
The region in blue corresponds to constraints from macroscopic-scale experiments that search for new forces \cite{Hoedl_MD_TP_2011,Terrano_MD_TP_2015,Youdin_MD_eN_1996,Ruoso_MD_eN_2017,Rong_MD_eN_2018}. 
The region in magenta corresponds to constraints from atomic and molecular electric dipole moment experiments \cite{Stadnik2017axion}.  
} 
\label{fig:Lin_S-PS_N-e}
\end{figure}

\begin{figure}[t!]
\includegraphics[width=\columnwidth]{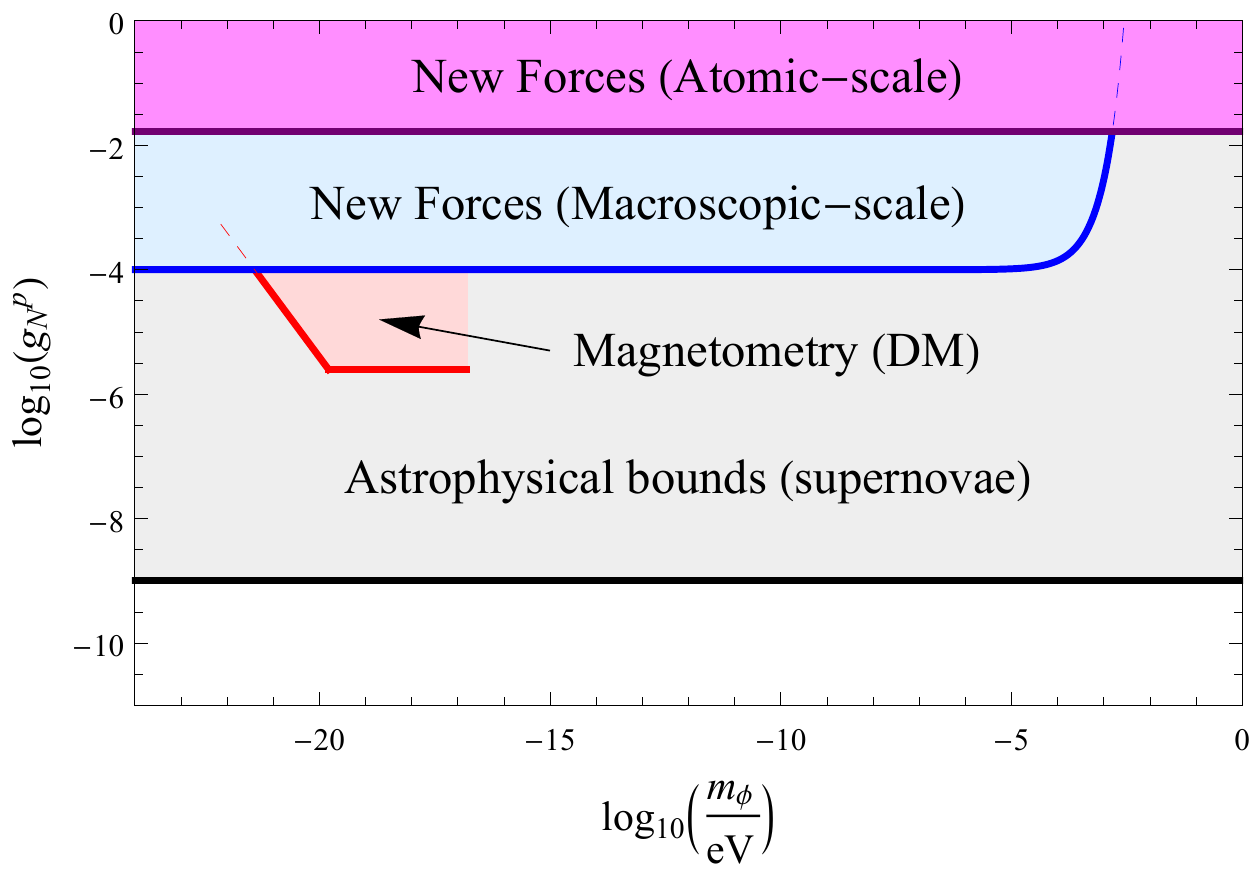}
\caption{Limits on the linear pseudoscalar interaction of a spinless boson $\phi$ with nucleons, as defined in the last term of Eq.~(\ref{axion_ints}). 
The region in blue corresponds to constraints from a macroscopic-scale experiment that searches for new forces \cite{Romalis_NF_2009}. 
The region in magenta corresponds to constraints from molecular hydrogen spectroscopy measurements and comparison with theory \cite{Ramsey_NF_1978}.  
The region in red corresponds to constraints from magnetometry measurements that search for the effects of a relic coherently oscillating field $\phi = \phi_0 \cos(m_\phi t)$, which saturates the local cold dark matter (DM) content \cite{nEDM_Axion_2017}. 
The region in light grey corresponds to astrophysical constraints pertaining to supernova energy-loss bounds \cite{Raffelt_AP_Review_1999,Raffelt_AP_Review_2008}. 
} 
\label{fig:Lin_pseudoscalar_nucleon}
\end{figure}

Current limits from macroscopic-scale experiments on several types of non-gravitational interactions of spinless bosons are shown in Figs.~\ref{fig:Lin_scalar_photon}, \ref{fig:Lin_S-PS_N-e} and \ref{fig:Lin_pseudoscalar_nucleon}.

\subsection{Atomic-scale experiments}
\label{Sec:Atomic_scale_exps}
Compared with the macroscopic-scale experiments discussed in Sec.~\ref{Sec:Macro_scale_exps}, the condition $m_\phi \ll 1/r$ holds up to much larger boson masses when a boson is exchanged between the constituents of an atom or molecule. 
This is because the interparticle separations between the constituents of atomic systems are much smaller than the length scales in macroscopic-scale experiments. 
Thus phenomena originating on atomic and sub-atomic length scales are generally much more sensitive to bosons with larger masses. 
Various atomic-scale phenomena can be used to search for new forces: 

\index{spectroscopy}
(1) Comparison of measured and predicted spectra of atoms, molecules and ions to search for new parity-conserving forces \cite{Ramsey_NF_1978,Karshenboim_NF_2010,Karshenboim_NF_2011,Kimball_NF_2013,Ficek_NF_2017,Fuchs_NF_2017,Ficek_NF_2018,Stadnik_2NE_2018}. 

\index{parity nonconservation}
(2) Comparison of measured and predicted parity-violating observables in atoms and molecules to search for new parity-violating forces \cite{Stadnik2017vector}. 

\index{electric dipole moment}
(3) Measurements of permanent electric dipole moments in atoms and molecules to search for new parity- and time-reversal-invariance-violating forces \cite{Stadnik2017axion,Stadnik2018axion}. 
\index{magnetometry}

There is an important difference between atomic-scale and macroscopic-scale experiments in the regime of a large boson mass, $m_\phi \gg 1/r$. 
Macroscopic-scale experiments lose sensitivity to new forces exponentially quickly when the boson mass becomes large, because the interaction becomes contact and so the very heavy boson cannot propagate between the source body and detector. 
In atomic-scale phenomena, however, there is always a finite probability for two constituent particles to be located very close to each other, and so these types of experiments lose sensitivity to new forces much more slowly (at a power-law rate) when the boson mass becomes large. 
Current limits from atomic-scale experiments on several types of non-gravitational interactions of spinless bosons are shown in Figs.~\ref{fig:Lin_S-PS_N-e} and \ref{fig:Lin_pseudoscalar_nucleon}.

\section{Laboratory Sources}
\label{Sec:Lab_sources}
In the presence of the non-gravitational interactions in the first terms of Eqs.~(\ref{lin_scalar_ints}) and (\ref{axion_ints}), spinless bosons may interconvert with photons. 
Several different types of methods can be used to exploit this possible interconversion:

\index{light-shining-through-a-wall}
(1) ``Light-shining-through-a-wall'' experiments \cite{Okun_LSW_1982,ALPS_2010,OSQAR_2015,GammeV_2008}. 
The basic idea here is to shine a powerful laser into a region of strong magnetic field. 
Some of the laser photons will convert into spinless bosons (provided that the energy of these photons is not less than the rest-mass energy of the spinless boson), which then pass through a wall that is impervious to photons (but not to the spinless bosons). 
A second strong magnetic field is applied on the other side of this wall, in order to reconvert some of the transmitted spinless bosons back into photons for detection. 
In principle, it is not necessary for all of the incident laser photons to be blocked by the wall. 
A tiny fraction of incident photons can be transmitted through the wall, so that an atomic or molecular transition can be resonantly induced involving the interference of photon- and spinless-boson-induced amplitudes (assuming there exists a non-gravitational interaction between the spinless boson and electron) \cite{Tran_LStW_2018}. 

\index{vacuum birefringence}
(2) Experiments to search for vacuum birefringence and dichroism \cite{Zavattini_BMV_1986,PVLAS_2016,QnA_2007,BMV_2008}. 
\index{dichroism}
The basic idea in these types of experiments is to shine a polarised laser into a region of strong magnetic field. 
Vacuum birefringence involves different indices of refraction for light polarised parallel and perpendicular to an applied magnetic field and is caused by virtual spinless bosons. 
Dichroism involves different absorptivities for light of different polarisations in an applied magnetic field and is caused by the production of real spinless bosons. 

\begin{figure}[t!]
\includegraphics[width=\columnwidth]{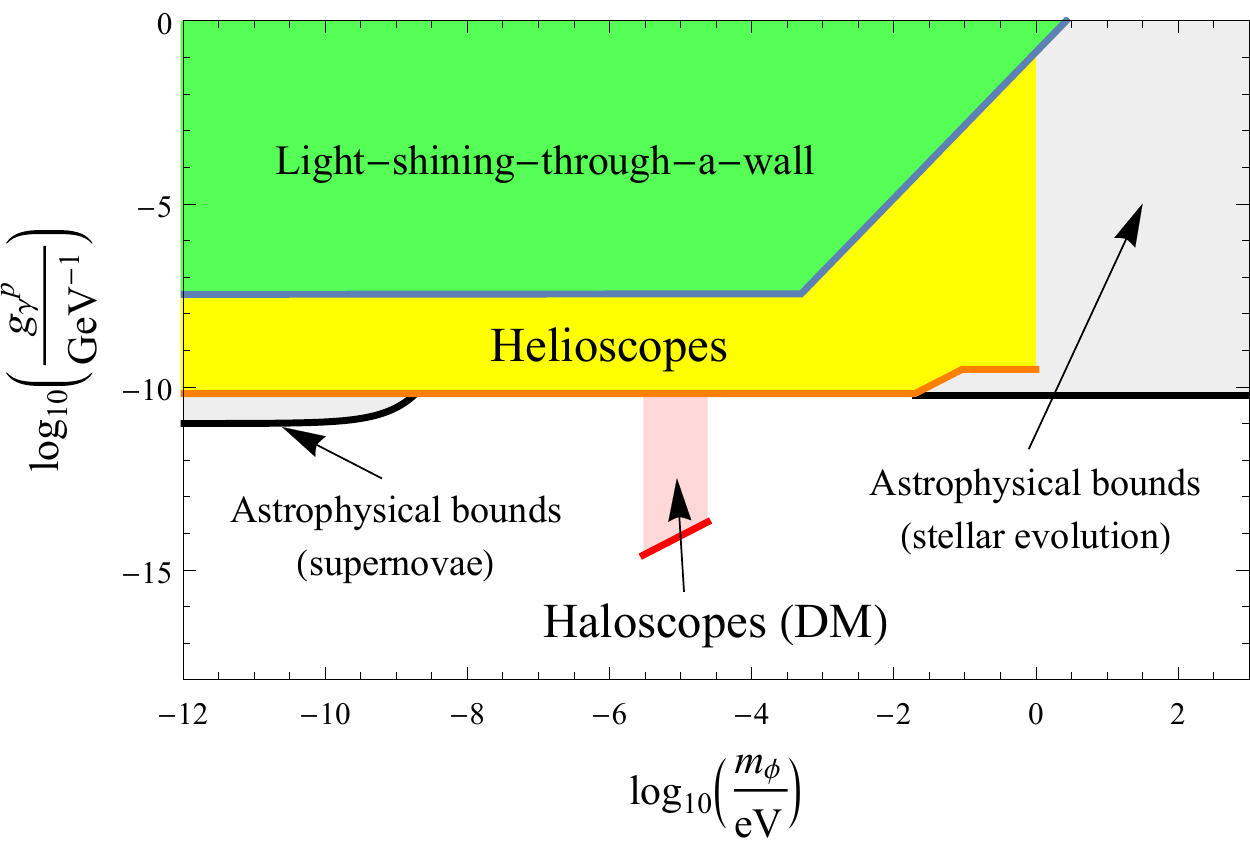}
\caption{Limits on the linear pseudoscalar interaction of a spinless boson $\phi$ with the photon, as defined in the first term of Eq.~(\ref{axion_ints}). 
The region in green corresponds to constraints from ``light-shining-through-a-wall'' experiments \cite{ALPS_2010,OSQAR_2015}. 
The region in yellow corresponds to constraints from helioscope experiments that search for bosons emitted from the Sun \cite{Sumico_2008,CAST_2017}. 
The region in red corresponds to constraints from haloscope experiments that search for the conversion of galactic dark matter (DM) bosons into photons \cite{ADMX_2017,ADMX_2018}. 
The regions in light grey correspond to astrophysical constraints pertaining to stellar evolution and supernova energy-loss bounds \cite{Raffelt_AP_Review_1999,Raffelt_AP_Review_2008}. 
} 
\label{fig:Lin_pseudoscalar_photon}
\end{figure}

Although electric fields could also be used in these experiments, in practice it is much easier to generate a stronger magnetic field (in terms of the equivalent electromagnetic energy density) in the laboratory. 
An advantage of producing spinless bosons in the laboratory is that the energies of these bosons are fixed by energy conservation, so resonance techniques can be applied without having to scan over an \emph{a priori} unknown range of boson energies. 
Current limits from laboratory source experiments on several types of non-gravitational interactions of spinless bosons are shown in Figs.~\ref{fig:Lin_scalar_photon} and \ref{fig:Lin_pseudoscalar_photon}.

\section{Astrophysical Sources}
\label{Sec:Astro_sources}
In the presence of the non-gravitational interactions in Eqs.~(\ref{lin_scalar_ints}), (\ref{quad_scalar_ints}) and (\ref{axion_ints}), spinless bosons can be produced and subsequently emitted from the hot interiors of active stars (such as the Sun) and dead stars (such as white dwarves), as well as in supernovae explosions. 
Excessive emission of spinless bosons from astrophysical sources would contradict observations and corresponding standard-model calculations, providing strong constraints on possible non-gravitational interactions of spinless bosons (see Figs.~\ref{fig:Lin_pseudoscalar_nucleon}, \ref{fig:Lin_pseudoscalar_photon}, \ref{fig:Lin_pseudoscalar_gluon} and \ref{fig:Quad_scalar_photon}) \cite{Raffelt_AP_Review_1999,Raffelt_AP_Review_2008}. 
If spinless bosons are emitted from the nearest star (the Sun), then it also becomes feasible to search for these particles in terrestrial experiments. 

\index{helioscope}
Spinless bosons emitted from the Sun can be detected using helioscope experiments, which seek to exploit the interconversion of spinless bosons with photons in a strong applied magnetic field \cite{Sikivie_Axion_1983,SOLAX_1998,Sumico_2008,CAST_2017,IAXO_2011}. 
Helioscope experiments are somewhat similar to the ``light-shining-through-a-wall'' experiments discussed in Sec.~\ref{Sec:Lab_sources}, except that the source of spinless bosons in helioscope experiments is provided by nature. 
The nature of the spinless boson sources in these two types of experiments is very different, however. 
In a ``light-shining-through-a-wall'' experiment, the energy of the spinless bosons depends on the frequency of the laser source used and can thus be altered. 
Additionally, since lasers are practically monochromatic sources of light, the resulting energy spectrum of spinless bosons is likewise sharply peaked in these types of laboratory experiments. 
On the other hand, in a helioscope experiment, the energy of the spinless boson is determined by the core temperature of the Sun ($\sim 1~\textrm{keV}$) and the energy spectrum of the bosons is relatively broad. 
Current limits from solar source experiments on one type of non-gravitational interaction of a spinless boson are shown in Fig.~\ref{fig:Lin_pseudoscalar_photon}.

\section{Cosmological Sources}
\label{Sec:Cosmo_sources}
\index{dark matter}
Low-mass (sub-eV) spinless bosons can be produced efficiently via non-thermal production mechanisms (which impart practically no kinetic energy to the bosons), such as ``vacuum misalignment'' in the early Universe \cite{Marsh2015Review}, and subsequently form a coherently oscillating classical field:~$\phi = \phi_0 \cos(\omega t)$, with the angular frequency of oscillation given by $\omega \approx m_\phi c^2 / \hbar$, where $m_\phi$ is the boson mass, $c$ is the speed of light and $\hbar$ is the reduced Planck constant. 
The classical nature of this field arises due to the large number of low-mass bosons per reduced de Broglie volume. 
The oscillating bosonic field carries the energy density $\rho_\phi \approx m_\phi^2 \phi_0^2 /2$, which may saturate the local cold dark matter (DM) energy density $\rho_\textrm{DM}^\textrm{local} \approx 0.4~\textrm{GeV/cm}^3$ \cite{Catena2010}. 
If these bosons comprise all of the DM, then the requirement that the boson de Broglie wavelength does not exceed the DM halo size of the smallest dwarf galaxies gives the lower boson mass bound $m_\phi \gtrsim 10^{-22}$ eV. 

A variety of atomic, molecular and optical experiments can be used to search for oscillating DM fields.  
The specific detection methods depend crucially on the particular non-gravitational interactions between the bosonic DM and ordinary matter that are probed. 
In this section, we give a brief overview of the main types of detection methods for oscillating bosonic fields. 
We note that similar detection strategies can also be implemented to search for bosonic fields that form ``clump-like'' DM, except in this case a network of detectors is required to unambiguously confirm the passage of such DM clumps \cite{Stadnik_TDM_2014,Pospelov_TDM_2014,Stadnik_Laser_DM_2015,Wcislo_TDM_2016,Roberts_TDM_2017}. 
\index{dark energy}
Additionally, spinless bosonic fields with certain self-interactions are conjectured in ``chameleonic'' models of dark energy \cite{Khoury_Chameleon_2004} and may be sought for with atom interferometry techniques \cite{Burrage_Chameleon_2015,Haslinger_Chameleon_2015}. 
\index{atom interferometry}

\subsection{Haloscope experiments}
\label{Sec:Haloscope_exps}
\index{haloscope}
In the presence of the non-gravitational interaction in the first term of Eq.~(\ref{axion_ints}), spinless bosons may interconvert with photons. 
Spinless bosons that make up (part of) the galactic DM can be detected using haloscope experiments, which aim to convert galactic DM bosons into photons in the presence of a strong applied magnetic field inside a microwave cavity \cite{Sikivie_Axion_1983,ADMX_2017,ORGAN_2017,ADMX_2018}. 
Haloscope experiments are examples of ``resonance-type'' experiments, since the resonant frequency of a cavity mode must match the boson's energy, $h \nu_\textrm{mode} \approx m_\phi c^2$. 
Although resonance-type experiments are sensitive to very feeble non-gravitational interactions, the drawback of these types of experiments is that the boson mass (and hence energy) are not known \emph{a priori}, meaning that these types of experiments have to scan over a very large range of frequencies in order to find a narrow resonance. 
Indeed, galactic bosonic DM in the vicinity of the Solar System is expected to have a root-mean-square velocity of $v_\textrm{rms} \sim 300~\textrm{km/s}$, giving the oscillating galactic bosonic field the finite coherence time:~$\tau_\textrm{coh} \sim 2\pi/m_\phi v_\textrm{rms}^2 \sim 2\pi \times 10^6 / m_\phi$, which is equivalent to a relative width of $\Delta \omega / \omega \sim 10^{-6}$. 
Current limits from haloscope experiments on one type of non-gravitational interaction of a spinless boson are shown in Fig.~\ref{fig:Lin_pseudoscalar_photon}.

 \begin{figure}[t!]
\includegraphics[width=\columnwidth]{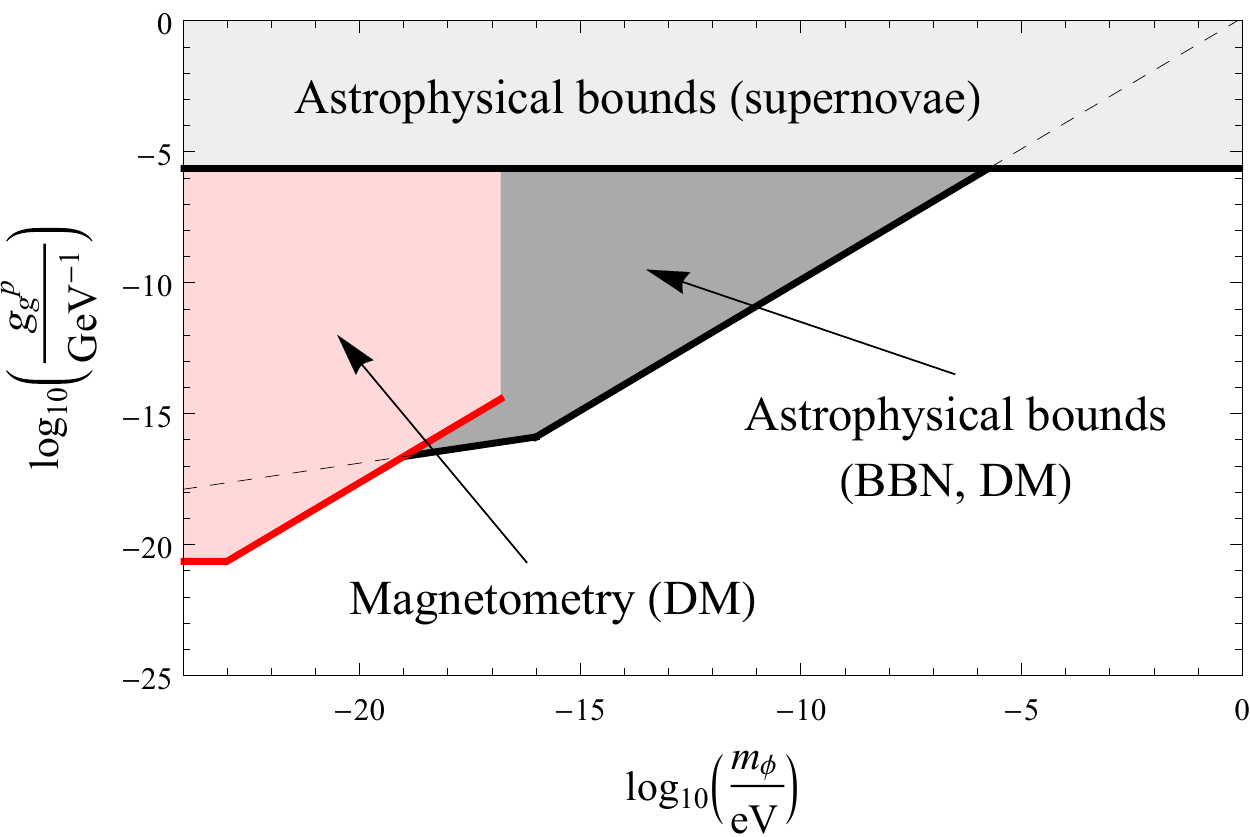}
\caption{Limits on the linear pseudoscalar interaction of a spinless boson $\phi$ with gluons, as defined in the second term of Eq.~(\ref{axion_ints}). 
The region in red corresponds to constraints from magnetometry measurements that search for the effects of a relic coherently oscillating field $\phi = \phi_0 \cos(m_\phi t)$, which saturates the local cold dark matter (DM) content \cite{nEDM_Axion_2017}. 
The region in light grey corresponds to astrophysical constraints pertaining to supernova energy-loss bounds \cite{Graham_Axion_2013}. 
The region in dark grey corresponds to astrophysical constraints pertaining to big bang nucleosynthesis (BBN) measurements, assuming that spinless bosons saturate the DM content \cite{Blum_Axion_BBN_2014,Stadnik_DM_VFC_2015,Stadnik_Thesis_2017}. 
} 
\label{fig:Lin_pseudoscalar_gluon}
\end{figure}

\subsection{Spin-precession experiments}
\label{Sec:Spin-precession_exps}
In the presence of the non-gravitational interactions in the last two terms of Eq.~(\ref{axion_ints}), bosonic DM fields can induce a number of time-varying spin-dependent effects. 
In particular, the second term in Eq.~(\ref{axion_ints}) gives rise to time-varying electric dipole moments of nucleons \cite{Graham_Axion_2011} and atoms and molecules \cite{Stadnik_Axion_2014}, with the angular frequency of oscillation governed by the boson mass. 
The last term in Eq.~(\ref{axion_ints}) gives rise to anomalous time-varying spin-precession effects due to the motion of Earth through an apparent time-varying \emph{pseudo}-magnetic field \cite{Flambaum_Axion_2013,Graham_Axion_2013,Stadnik_Axion_2014}. 
Various types of methods can be used to search for time-varying spin-dependent effects: 

\index{magnetometry}
(1) Atomic magnetometers and ultracold neutrons to search for time-varying anomalous spin-precession effects \cite{Flambaum_Axion_2013,Stadnik_Axion_2014,Stadnik_Thesis_2017,nEDM_Axion_2017}. 
\index{electric dipole moment}

\index{torsion pendulum}
(2) Torsion pendula to search for time-varying anomalous torques \cite{Flambaum_Axion_2013,Stadnik_Axion_2014,Stadnik_Thesis_2017}. 

(3) Nuclear-magnetic-resonance techniques to search for the resonant build-up of transverse magnetisation \cite{Graham_Axion_2013,CASPEr_2014,CASPEr_2018}. 

(4) Resonant conversion of galactic DM bosons into photons in a magnetised material \cite{Krauss_Axion_1985,Barbieri_Axion_1989,Kakhidze_Axion_1991,QUAX_2017}. 

In the case of time-varying electric dipole moments, which can be sought for with methods (1) and (3), it is necessary to apply an electric field in the experiment. 
In the case of methods (1), (2) and (3), the observables scale only to the first power of the underlying interaction constant. 
This is a particularly attractive feature of these types of methods, compared with the methods discussed in Secs.~\ref{Sec:New_forces}, \ref{Sec:Lab_sources}, \ref{Sec:Astro_sources} and \ref{Sec:Haloscope_exps}, where the observables scale either to the second or fourth power in a small interaction constant. 
Current limits from spin-precession experiments on several types of non-gravitational interactions of spinless bosons are shown in Figs.~\ref{fig:Lin_pseudoscalar_nucleon} and \ref{fig:Lin_pseudoscalar_gluon}.

 \begin{figure}[t!]
\includegraphics[width=\columnwidth]{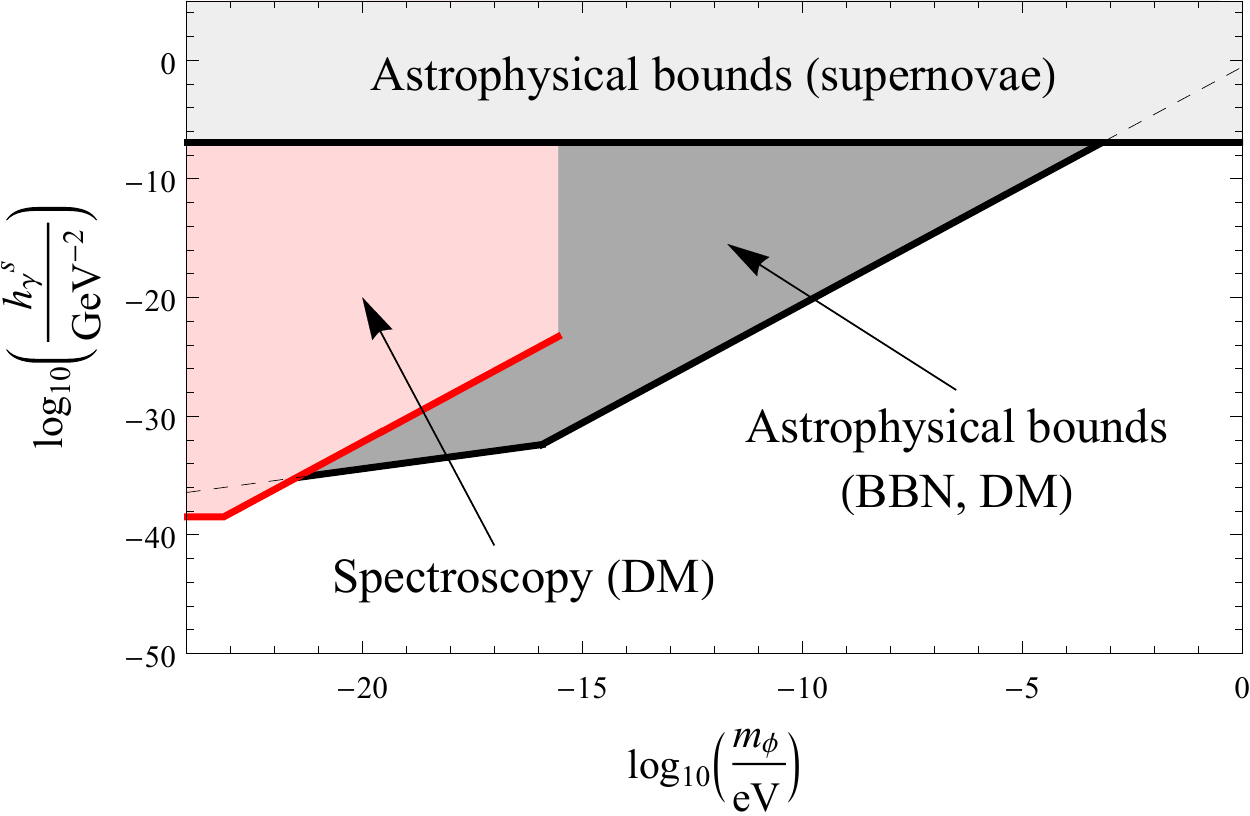}
\caption{Limits on the quadratic scalar interaction of a spinless boson $\phi$ with the photon, as defined in the first term of Eq.~(\ref{quad_scalar_ints}). 
The region in red corresponds to constraints from atomic spectroscopy measurements that search for the effects of a relic coherently oscillating field $\phi = \phi_0 \cos(m_\phi t)$, which saturates the local cold dark matter (DM) content \cite{Stadnik_DM_VFC_2015,Stadnik_DM_VFC_2016}. 
The region in light grey corresponds to astrophysical constraints pertaining to supernova energy-loss bounds \cite{Pospelov_2BE_2008}. 
The region in dark grey corresponds to astrophysical constraints pertaining to big bang nucleosynthesis (BBN) measurements, assuming that spinless bosons saturate the DM content \cite{Stadnik_DM_VFC_2015}. 
} 
\label{fig:Quad_scalar_photon}
\end{figure}

\subsection{Time-varying physical constants}
\label{Sec:Time-varying_constants}
In the presence of the non-gravitational interactions in Eqs.~(\ref{lin_scalar_ints}) and (\ref{quad_scalar_ints}), bosonic DM fields can induce ``apparent'' variations in the physical constants \cite{Stadnik_Laser_DM_2015,Tilburg_DM_VFC_2015,Stadnik_DM_VFC_2015}. 
\index{spectroscopy}
One particularly powerful class of measurements to search for these apparent oscillations in the physical constants involve high-precision comparisons of atomic and molecular transition frequencies \cite{Stadnik_Laser_DM_2015,Tilburg_DM_VFC_2015,Stadnik_DM_VFC_2015,Leefer_DM_VFC_2015,Hees_DM_VFC_2016,Stadnik_DM_VFC_2016}, which have previously been used to search for ``slow temporal drifts'' in the physical constants \cite{Dzuba_VFC_Review_2009,Chapter30}. 
\index{atomic clock}
The basic idea of clock-comparison experiments is to use two transition frequencies with different sensitivities to variations in the physical constants. 
For example, in the atomic units $\hbar = e = m_e = 1$, an atomic optical transition frequency scales as $\omega_\textrm{opt} \propto F^\textrm{opt}_\textrm{rel}(Z \alpha)$, while an atomic hyperfine transition frequency scales as $\omega_\textrm{hf} \propto [\alpha^2 F^\textrm{hf}_\textrm{rel}(Z \alpha)] (m_e/m_N) \mu$, where $F_\textrm{rel}$ are relativistic factors (which generally increase rapidly with the nuclear charge $Z$), and $\mu$ is the nuclear magnetic dipole moment. 
A summary of calculated sensitivity coefficients for various atomic, molecular and nuclear transitions can be found in \cite{Dzuba_VFC_Review_2009,Chapter30}.

\index{laser interferometry}
Instead of comparing two transition frequencies, it is also possible to compare a transition frequency against a reference frequency determined by the length of an optical cavity or an interferometer arm \cite{Stadnik_Laser_DM_2015,Stadnik_Cavity_DM_2016}. 
In this case, the reference frequency scales roughly as $\omega_\textrm{ref} \propto 1/L \propto 1/a_\textrm{B}$. 
The sensitivity coefficients for laser/maser interferometry experiments depend on the specific mode of operation and have been calculated in \cite{Stadnik_Laser_DM_2015,Stadnik_Cavity_DM_2016}. 
Like some of the methods discussed in Sec.~\ref{Sec:Spin-precession_exps} to search for time-varying spin-dependent effects, the observables in the types of experiments discussed in this section also have the attractive feature of scaling to the first power of the underlying interaction constant. 
Current limits on several types of non-gravitational interactions of spinless bosons from experiments that search for apparent oscillations in the physical constants are shown in Figs.~\ref{fig:Lin_scalar_photon} and \ref{fig:Quad_scalar_photon}.

\pagebreak
\section{Biographies}

{\em
\noindent{\bf Victor Flambaum\\
University of New South Wales\\
Department of Physics\\
Sydney, Australia\\
v.flambaum@unsw.edu.au\\ \/}
Prof.~Dr.~Victor Flambaum is a Professor of Physics and holds a Chair of Theoretical Physics. 
Ph.D., D.Sc.~from the Institute of Nuclear Physics, Novosibirsk, Russia. 
He has about 400 publications in atomic, nuclear, particle, solid state, statistical physics, general relativity and astrophysics including works on violation of fundamental symmetries (parity, time reversal, Lorentz), test of unification theories, temporal and spatial variation of fundamental constants from Big Bang to present, many-body theory and high precision atomic calculations, as well as statistical theory of finite chaotic Fermi systems and enhancement of weak interactions.\\
\/}

\:
\:
\:
\:
\:

{\em
\noindent{\bf Yevgeny Stadnik\\
Johannes Gutenberg University of Mainz\\
Helmholtz Institute Mainz\\
Mainz, Germany\\
stadnik@uni-mainz.de\\ \/}
Dr.~Yevgeny Stadnik is a Humboldt Research Fellow at the Johannes Gutenberg University of Mainz. 
He completed his Ph.D.~at the University of New South Wales, Sydney, Australia. 
His research interests include the manifestations of dark matter and new particles in low-energy atomic and astrophysical phenomena, as well as tests of the electroweak theory.\\
\/}

\onecolumn{
\small{

}
}

\printindex

\end{document}